\documentclass[aps, preprint,showpacs,preprintnumbers,amsmath,amssymb]{revtex4-1}
\usepackage[colorlinks=true, pdfstartview=FitV, linkcolor=red, citecolor=blue, urlcolor=black, pdftitle={},pdfauthor={Yuki Minami, Yoshimasa Hidaka},pdfsubject={}, pdfkeywords={Derivative expansion, causality, relativistic theory, effective theory}]{hyperref}

\usepackage{amsmath,amssymb,bm}
\usepackage[dvips]{graphicx}

\newcommand{\G}{G_R}

\begin{document}
\preprint{RIKEN-MP-81, RIKEN-QHP-104}

\author{Yuki Minami}
\affiliation{Theoretical Research Division, Nishina Center, RIKEN, 
             Wako 351-0198, Japan}
\author{Yoshimasa Hidaka}
\affiliation{Theoretical Research Division, Nishina Center, RIKEN, 
             Wako 351-0198, Japan}  

\title{General conditions ensuring  relativistic causality in an effective field theory based on the derivative expansion}
\begin{abstract}
We discuss the general conditions ensuring relativistic causality in an effective theory  based on the derivative expansion.
Relativistic causality implies that the Green function vanishes in the space-like region.
It is known that a naive derivative expansion violates causality in some cases such as the first-order relativistic dissipative hydrodynamics.
We note that the Lorentz covariance, and the equal order of time and space derivatives do not ensure causality.
We derive the general conditions for causality that should be satisfied by any effective theories respecting special relativity.
The conditions are the followings:
(i) the imaginary part of poles of the Green function is bounded at the large momentum limit, 
(ii) the front velocity is  smaller than the speed of light, and
(iii) the coefficient of the highest-order time derivative does not include space derivatives. 
\end{abstract}	

\maketitle

\section{Introduction}
Derivative expansion is a useful tool at low-energy scale and widely used in effective theories.
Chiral perturbation theory is one of great successful examples as the low-energy effective theory in hadron physics~\cite{weinberg, Weinberg1979327, Ecker:1994gg}.
In modern point of view, the hydrodynamics is also a low-enegy effective theory; the leading-order hydrodynamic equations are called Euler equations, and  the first order of hydrodynamic equations are called Navier-Stokes equations.

Causality is an important concept in physics. 
For relativistic systems, the propagation of any information cannot exceed the speed of light (relativistic causality).
However, it seems that a low-energy effective theory  in medium is incompatible with relativistic causality.
For example,  in the first-order relativistic hydrodynamics, the shear and heat flows violate causality because it has the form of a diffusion equation~\cite{Israel:1976tn,Israel:1979wp, Pu:2009fj, Koide:2011tj, Denicol:2008ha, Buchel:2009tt, Baier:2007ix}.

One might regard acausality in the effective theory as 
 a problem about range of  validity~\cite{geroch, Lindblom19961, Kostadt:2000ty, *Kostadt:2001rr, Van:2007pw}. 
 The low-energy effective theory based on the derivative expansion has an ultraviolet cutoff that separates micro- and macro-scopic degrees of freedom.
One may think that the violation of causality is not the problem as long as the violation is much smaller than the cutoff scale.
However, such a small violation will be amplified by Lorentz boost, and it exceeds the UV cutoff scale~\cite{Hiscock:1985zz,Denicol:2008ha}.


One also might expect  that acausality in the diffusion equation is caused by  non-Lorentz covariance of the equation.   
However, the Lorentz covariance does not ensure causality. 
In fact, the relativistic hydrodynamic equations in the first order of the derivative expansion violate causality even though the equations are covariant~\cite{Israel:1976tn,Israel:1979wp, Pu:2009fj, Koide:2011tj, Denicol:2008ha, Buchel:2009tt,Baier:2007ix}.
For a concrete example, let us consider the Lorentz covariant diffusion equation,
\begin{align}
\biggl[ u^\mu \partial_\mu + \Gamma (\eta^{\mu \nu}-u^\mu u^\nu )\partial_\mu \partial_\nu \biggr]n(x^\mu)=0,
 \label{eq:covariantdiffusion}
\end{align} 
where $u^\mu$ is a time-like vector satisfying $u^2=1$, $n(x^\mu)$ a scalar density, $\Gamma$ the diffusion constant, $\eta^{\mu \nu} = \mathrm{diag}  (1, -1, -1, -1) $ the Minkowski metric. 
If we choose $u^\mu = (1, \bf{0})$, Eq.~(\ref{eq:covariantdiffusion}) becomes the ordinary diffusion equation, $\bigl( \partial_0  - \Gamma \partial_i^2 \bigr) n(x^\mu) =0$.
The retarded Green function of Eq.~(\ref{eq:covariantdiffusion})  for constant $u^\mu$ becomes
\begin{align}
\G(x^\mu)=\theta(x_t) \biggl( \frac{1}{\pi \Gamma x_t} \biggr)^{3/2}\exp\biggl[ \frac{x_s^2}{4\Gamma x_t }\biggr] ,
\end{align}
where $\theta(x_t)$ is the step function,  $x_t \equiv u_\mu x^\mu$ and $x_s^\mu \equiv (\eta^{\mu \nu} - u^\mu u^\nu) x_\mu$ are 
the time- and space-like components of $x^\mu$, respectively.  
We can see that the retarded Green function does not vanish in the space-like region, $x^2=x_t^2 + x_s^2 < 0$.
Therefore, the Lorentz covariance of the equation is not sufficient to ensure causality.

It is argued that acausality comes from the difference between the order of  time- and space-like derivatives in the equation of motion~\cite{Israel:1976tn, Israel:1979wp, Pu:2009fj, Denicol:2008ha}.
In the diffusion equation Eq.~(\ref{eq:covariantdiffusion}), the time-like derivative is the first order while the space-like one is the second order.
This different order is cured by introducing $\tau ( u^\mu \partial_\mu )^2$ to the equation:
\begin{align}
\biggl[ \tau ( u^\mu \partial_\mu )^2  +u^\mu \partial_\mu 
       + \Gamma (\eta^{\mu \nu}-u^\mu u^\nu)\partial_\mu\partial_\nu \biggr]n(x^\mu)=0,
\end{align}
where $\tau$ is a parameter corresponding to the relaxation time. 
This equation has the form of the telegraphic equation, so that the propagation is restricted in the region, $v x_t  >   x_s $, where $v=\sqrt{\Gamma/\tau}$ and $x_s = \sqrt{-x_s^2}$.
If $\Gamma<\tau$,  causality is satisfied because the velocity is smaller than the speed of right, $v<1$~\cite{Pu:2009fj, Denicol:2008ha}.
However, if $\Gamma>\tau$, the propagation speed exceeds the speed of light. 
Furthermore, if $\Gamma=0$, causality is not violated even though the order of the time and spatial derives are different.
Therefore, the equal order of space and time derivatives in the equation of motion itself does not mean that the Green function is causal.

We note that the group velocity with a finite wavenumber is not good quantity for discussing causality.
The group velocity is given as
\begin{align}
v_g (k) \equiv \frac{\partial\, \mathrm{Re} \,\omega(k)}{\partial k},
\end{align}  
where $ k = |\bm{k}|$ is the momentum, and $\omega(k)$ is the pole of the retarded Green function.
It is known that the retarded Green function can vanish in the space-like region even if $v_g (k)$ is faster than the speed of light at some momentum.
For $\omega (k)\propto k$ at $k \rightarrow \infty$, it is shown that causality is satisfied 
if the front velocity $v_f \equiv \lim_{k \rightarrow \infty} v_g (k)$ is slower than the speed of light~\cite{Brillouin, milonni2004fast, Chaio, Pu:2009fj, Koide:2011tj}.
This was firstly pointed out by Sommerfeld and Brilliouin in the context of light pulse propagation in medium~\cite{Brillouin}.
They considered propagation of discontinuous wave front to define the information propagation, which is described by the front velocity.
The group velocity describes the propagation of the peak of the pulse, but does not determine the information propagation~\cite{milonni2004fast, Chaio}. 
We note that this condition for the front velocity does not cover acausality for the diffusion mode, whose pole is 
\begin{align}
\omega_{\text{diff}} (k) &\propto -i k^2. \label{eq:diffusionmode}
\end{align}
We see that causality is violated although the front velocity of diffusion mode vanishes., i.e., smaller than the speed of light. 

What ensures causality in general?
In  quantum field theory,  relativistic causality is ensured by
\begin{equation}
[\phi(x), \phi(y)] = 0 \;\; \text{for}  \;\; x^2-y^2 <0,
\end{equation}  
where $\phi(x)$ is a Heisenberg operator, $[.. , ...]$ the commutation relation.
The retarded propagator also vanishes in the space-like region:
\begin{align}
\G(x^\mu-y^\mu) &\equiv \theta( x^0 - y^0 )\langle [\phi(x), \phi(y)] \rangle, \nonumber \\
         &=0 \;\; \text{for} \;\; x^2-y^2<0.  \label{eq:causality}
\end{align} 
The low-energy poles of the propagator correspond to those of the Green function in the low-energy effective theory.
Therefore, Eq.~(\ref{eq:causality}) should be satisfied even in the low-energy Green function if the theory respects special relativity.

The purpose of this paper is to derive the conditions ensuring relativistic causality in an effective theory based on the derivative expansion.
We will consider the retarded Green function in a scalar theory at tree level, i.e., thermal and quantum fluctuation will  not be taken into account.
In this case,  the retarded Green function in the derivative expansion is generally written as a rational function in the momentum space:
\begin{equation}
\G(\omega, k ) = \frac{Q(\omega, k)}{P(\omega, k)}, \label{eq:Gpoly}  
\end{equation}   
where $P(\omega, k) $ and $Q(\omega , k)$ are polynomials in $\omega $ and $k$:
\begin{align}
P(\omega , k) =  p_n (k) \omega^n + p_{n-1}(k)\omega^{n-1} + ... + p_1(k) \omega +p_0(k), \\
Q(\omega, k) =   q_m (k)\omega^m + q_{m-1}(k)\omega^{n-1} + ... + q_1(k) \omega +q_0(k).
\end{align}
Here, $n>m$, and $p_j (k)$ and $q_j(k)$ are the polynomials in $k$.
 We assumed isotropy, and then the Green function turns out to be a function of $k$.
We will discuss relativistic causality based on Eqs.~(\ref{eq:causality}) and (\ref{eq:Gpoly}). 

In the following sections, we will derive the general conditions ensuring that the retarded Green function vanishes in the space-like region,
which are given by
\begin{align}
{\lim_{k \to \infty}\biggl| \mathrm{Re}\,}\frac{\omega(k)}{k} \biggr| &< 1,\label{eq:recondition}\\
 \lim_{k \to \infty} \biggl|\mathrm{Im}\frac{\omega(k)}{k} \biggr| &< \infty , \label{eq:imcondition}
\end{align}
and that $p_n (k)$ must not depend on $k$.
These conditions ensure causality in effective theories  based on the derivative expansion, and are our main results in this paper. 
We note that the first condition, Eq.~(\ref{eq:recondition}), is nothing but the condition that the front velocity is smaller than the speed of light because,  if $\omega (k)\propto k$ at larger $k$,  we have 
\begin{align}
\lim_{k\to\infty} \left| \frac{{\mathrm{Re}\,} \omega (k) }{k}\right| = \lim_{k\to\infty}\left|\frac{\partial\, { \mathrm{Re}\,} \omega (k)}{\partial k} \right|= v_f <1.
\end{align}

This paper is organized as follows. In Sec.~\ref{sec:Derivation}, we derive the general conditions to ensure causality in an effective theory based on the derivative expansion using the retarded Green function.
In Secs.~\ref{sec:ContributionFromBranchCuts} and \ref{sec:ContributionFromPole}, we evaluate the contributions from branch cuts and poles to the Green function, respectively. Section~\ref{Sec:Summary} is devoted to summary.

\section{Derivation of the general conditions ensuring causality} \label{sec:Derivation}
As mentioned in the previous section, relativistic causality implies that the propagation vanishes for the space-like region, i.e.,
$\G(x_\mu)=0$ for $x^2<0$. In this section, we show that this holds if Eqs.~(\ref{eq:recondition}) and (\ref{eq:imcondition}) are satisfied.

Let us start with the retarded Green function in momentum space.
In an effective theory base on the derivative expansion, the retarded Green function is given as the rational function Eq.~(\ref{eq:Gpoly}).
In order to discuss causality, we employ the partial-fraction decomposition for  the retarded Green function,
\begin{equation}
\G (\omega, k ) = \sum_{i, j} \frac{f_{i, j} (k)}{\big(\omega -\omega_j(k)\big)^i }, 
\end{equation}
where $\omega_j(k)$ denote the position of poles on the complex $\omega$-plane.
Since we consider the retarded Green function, the poles are located on the lower half-plane, i.e., $\mathrm{Im}\,\omega_j(k)<0$.
We consider the case that  orders of all the poles are one.
The higher-order poles can be treated as first-order poles by infinitesimally splitting the pole positions.
For example,  a second-order pole is rewritten as
\begin{align}
\frac{1}{(\omega-\omega_j(k))^2} 
                                   =\lim_{\epsilon \rightarrow 0} \frac{1}{2\epsilon} 
                                     \biggl( \frac{1}{\omega-\omega_j(k)-\epsilon}- \frac{1}{\omega-\omega_j(k)+\epsilon }\biggr). \label{eq:second-order-pole}
\end{align}
Therefore, the following argument is valid if the $\epsilon\to 0$ limit can be smoothly taken after calculations.
In this case, the retarded Green function is written as
\begin{equation}
\G (\omega, k ) = \sum_{j} \frac{f_{j} (k)}{\omega -\omega_j(k) }, \label{eq:Gpf}
\end{equation}
with 
\begin{equation}
f_j(k) =\lim_{\omega \to \omega_j(k)}\biggl[ \big(\omega - \omega_j(k)\big)\frac{Q(\omega,k)}{P(\omega, k)} \biggr]. \label{eq:fjk}
\end{equation}
The retarded Green function in coordinate space is given by the Fourier transformation of $\G(\omega, k)$:
 \begin{equation}
 \G (x^\mu) = \int \frac{d \omega}{2\pi}\frac{d^3k}{(2\pi)^3} G(\omega, k) e^{-i \omega t + i \bm{k} \cdot \bm{x}}.\\
  \end{equation} 
This integral becomes zero for $t<0$ because the pole is located on the lower half-plane. For $t>0$, the retarded Green function becomes
 \begin{align}
 \G (x^\mu)  &= \int  \frac{d \omega}{2\pi}\frac{d^3k}{(2\pi)^3} 
                    \sum_j \frac{f_j (k)}{\omega -\omega_j(k) } e^{-i \omega t + i \bm{k} \cdot \bm{x}},  \notag\\
               &= i \sum_j \int\frac{d^3k}{(2\pi)^3} f_j (k) e^{-i \omega_j(k) t + i \bm{k} \cdot \bm{x}}, \notag\\
               &=\frac{1}{4\pi^2 r} \sum_j \int_{-\infty}^{\infty}dk k f_j(k) \exp \biggl[ ik \biggl( r -\frac{\omega_j(k)}{k}t  \biggr)\biggr],
\label{eq:G}
  \end{align} 
where $r\equiv |\bf{x}|$.

\begin{figure}
\centering
  \begin{tabular}{cc}
    \includegraphics[width=0.45\hsize]{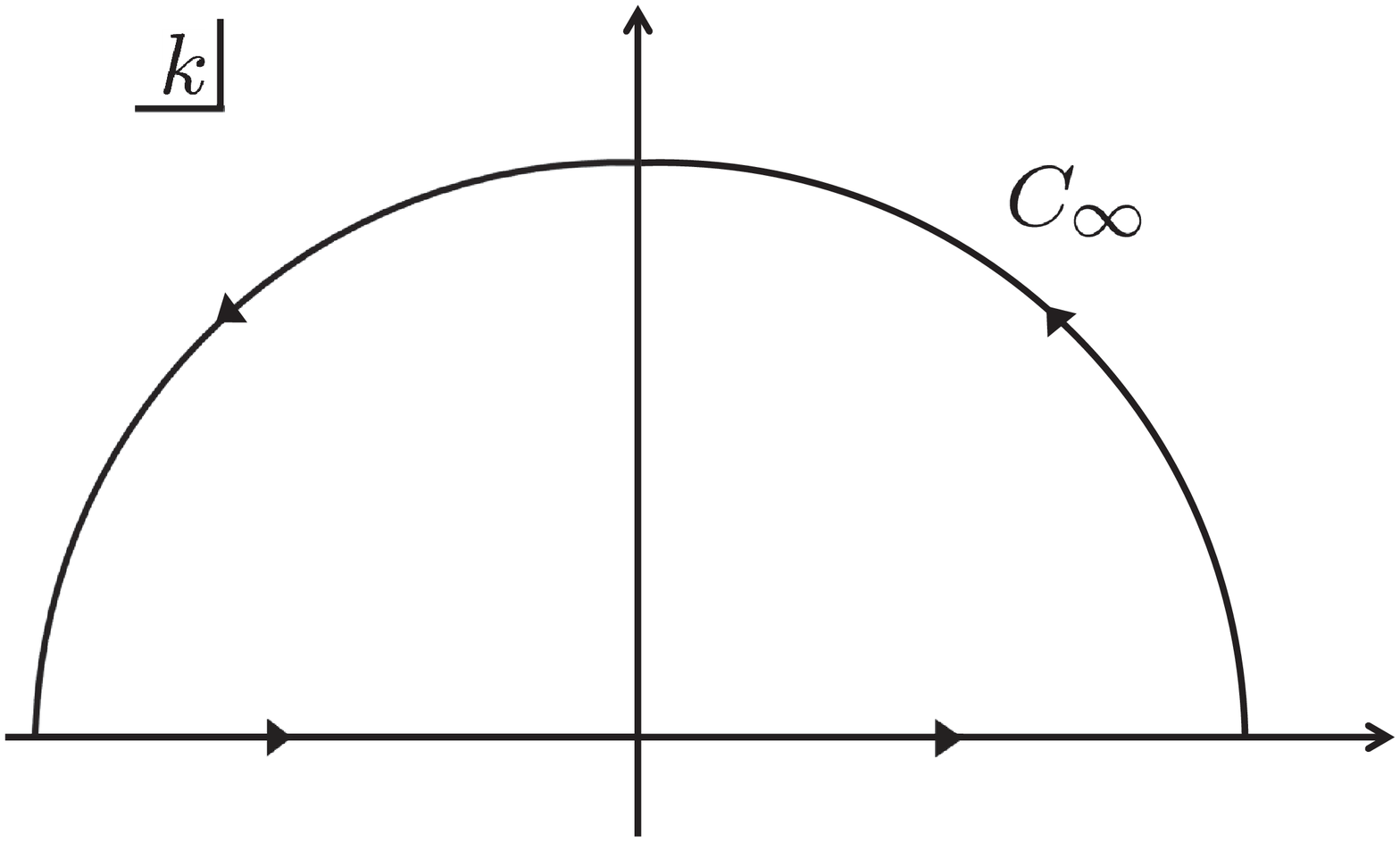}&
    \includegraphics[width=0.45\hsize]{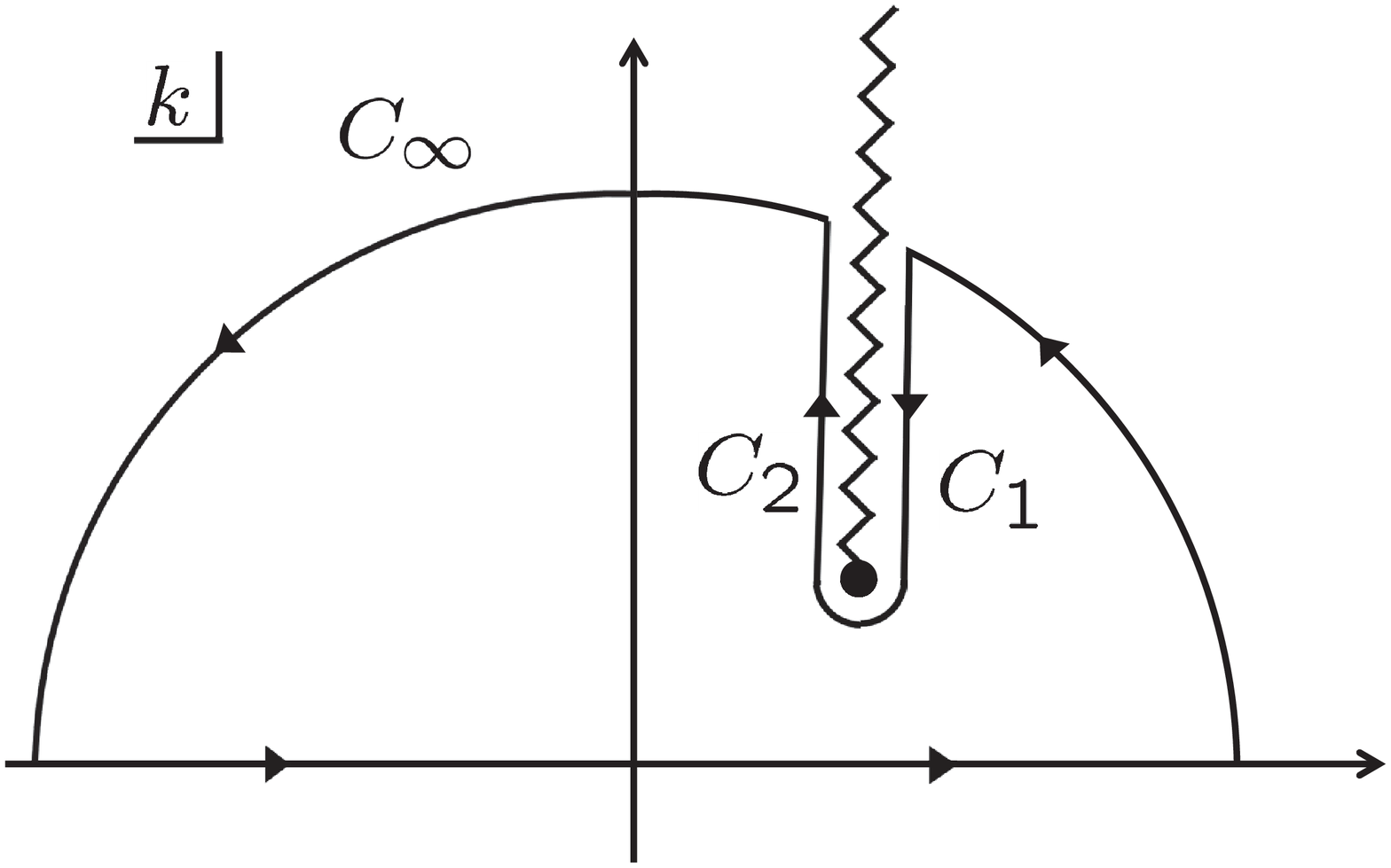}\\
    (a)& (b)
    \label{fig:contour}
    \end{tabular}
  \caption{The contours for evaluating the integral Eq.~(\ref{eq:G})  without  branch cuts (a) and 
  with a branch cut (b).}
\end{figure}

Now, we discuss the conditions ensuring that  the retarded Green function vanishes in the space-like region, $ r - t >0$.
To evaluate the integral with respect to $k$, we consider the complex integral on the complex $k$-plane.
If the integrand does not have any poles and branch cuts, 
we can evaluate the integral along the contour in Fig.~\ref{fig:contour}a.
The retarded Green function is equal to the contribution from $C_{\infty}$,
which vanishes if $\omega_j(k)$ satisfies Eqs.~(\ref{eq:recondition}) and (\ref{eq:imcondition}).
If this is not the case, $ r -(\omega_j(k) / k) t $ changes the sign at some point on $C_{\infty}$, 
and the contribution from $C_{\infty}$ does not generally vanish.
Therefore, Eqs.~(\ref{eq:recondition}) and (\ref{eq:imcondition}) must be satisfied to ensure causality.
 
In general, $f_j(k)$ may contain cuts and poles on the complex $k$-plane. 
In the following subsections, 
we evaluate contributions from these poles and cuts to the integral, and show that they cancel out 
if $p_n (k)$ does not depends on $k$. 
Furthermore, we discuss Eq.~(\ref{eq:G}) cannot vanish in the space-like region if $p_n (k)$ depends on $k$.

\subsection{Contributions from  branch cuts}\label{sec:ContributionFromBranchCuts}
Here, we evaluate the contributions from  branch cuts.
First, let us factorize the denominator of the propagator in Eq.~(\ref{eq:Gpoly}) into the $\omega$-independent part and others,
\begin{equation}
P(\omega, k) =p_n (k) \prod_j^n (\omega -\omega_j(k) ), \label{eq:denominator}
\end{equation}
where  $\omega_j(k)$ may have branch cuts on the complex $k$-plane, while $p_n (k)$ is analytic.
We consider a simple situation that $\omega_j(k)$ have only one branch cut on the upper half-plane,  like Fig.~\ref{fig:contour}a.
The generalization to the case that $\omega_j(k)$ have multi-branch cuts is straightforward.

We suppose that $\omega_{j_1}$ has a cut. If  we pass across the cut, we have the discontinuity,
\begin{equation}
\lim_{\epsilon\to0}\omega_{j_1} (s+ \epsilon ) \neq \lim_{\epsilon\to0}\omega_{j_1} (s - \epsilon),
\end{equation}
where $s$ is a point on the cut.
In contrast,  $P(\omega, k) $ does not have the discontinuity on the cut,
\begin{equation}
\lim_{\epsilon\to0}P(\omega, s+\epsilon) =\lim_{\epsilon\to0} P(\omega, s-\epsilon),
\end{equation} 
because the $P(\omega,k)$ is polynomial and analytic on complex $k$-plane.
This implies that $\omega_{j_1}$ has the pair $\omega_{j_2}$ such that
\begin{equation}
\lim_{\epsilon\to0}\omega_{j_1} (s+ \epsilon ) = \lim_{\epsilon\to0}\omega_{j_2} (s - \epsilon), \label{eq:pair}
\end{equation}
for $j_1 \neq j_2$.

Next, let us evaluate the integral Eq.~(\ref{eq:G}), which can be performed by the integral along the contour in Fig.~\ref{fig:contour}b.
Noting that  the residue  $f_{j_{1,2}} (k)$ has the same branch cut as that of $\omega_{j_{1,2}}$ [See Eq.~(\ref{eq:fjk})],
we have 
\begin{align}
\G(x^\mu) &= \frac{1}{4\pi^2 r} \sum_j  \int_{C_1+C_2} dk k f_j (k) e^{-i \omega_j(k)t+ i k r}, \notag \\
             &= \frac{1}{4\pi^2 r}  \int_{C_1+C_2} dk k\bigg[  f_{j_1} (k) e^{-i \omega_{j_1}(k)t+ i k r}
                      +f_{j_2} (k) e^{-i \omega_{j_2}(k)t+ i k r}\biggr]. \label{eq:cut12} 
\end{align}
Here, we assumed that Eqs.~(\ref{eq:recondition}) and (\ref{eq:imcondition}) are  satisfied, and dropped the contribution from $C_\infty$. 
From Eqs. (\ref{eq:fjk}) and (\ref{eq:pair}), we obtain the following relations:
\begin{align}
\int_{C_2} f_{j_1} (k) e^{-i \omega_{j_1}(k)t+ i k r} &= - \int_{C_1} f_{j_2} (k) e^{-i \omega_{j_2}(k)t+ i k r}, \\
\int_{C_2} f_{j_2} (k) e^{-i \omega_{j_2}(k)t+ i k r} &= - \int_{C_1} f_{j_1} (k) e^{-i \omega_{j_1}(k)t+ i k r}.                              
\end{align}
Then, the first and second terms in the last line of  Eq.~(\ref{eq:cut12}) cancel out.
Therefore, the conditions, Eqs.~(\ref{eq:recondition}) and (\ref{eq:imcondition}), does not change even if the branch cuts exist.

\subsection{Contributions from poles}\label{sec:ContributionFromPole}
Next, we discuss the poles of $f_j(k)$.
From Eq.~(\ref{eq:fjk}), $f_j (k)$ can be written as
\begin{align}
f_j(k) = \frac{Q(\omega_j(k),k)}{p_n (k) \prod_{i \neq j} (\omega_j (k)- \omega_i (k))}.
\end{align}
The poles on the complex $k$-plane are given as the solution of the followings:
\begin{align}
p_n(k) &= 0, \label{eq:pnkpole}  \\
\quad \omega_j (k)- \omega_i (k) &=0. \label{eq:omegapole}
\end{align}
We will show that poles from Eq.~(\ref{eq:omegapole}) does not contribute to Eq.~(\ref{eq:G}),
whereas the contribution of those from Eq.~(\ref{eq:pnkpole})  does not vanish and violates causality.
Thus, $p_n(k)$, which is the coefficient of the highest-order time derivative, must not depend on $k$ for causality.

First, we show that contribution of  poles from Eq.~(\ref{eq:omegapole}) cancels out. 
We here set $p_n(k) \rightarrow p_n$.
We suppose that  a pair, $\omega_{j_1}(k) $ and $ \omega_{j_2}(k)$, satisfies  $\omega_{j_1} (k)- \omega_{j_2} (k) =0$ at $k=k_c$ on the upper half-plane, and the solution is the first order.
We also assume that $\omega_{i} (k)- \omega_{j} (k) \neq 0$ if $i,j\neq j_1$, $j_2$. 
Thus, we can write the difference of $\omega_{j_1}(k)$ and $\omega_{j_2}(k)$ as
\begin{equation}
\omega_{j_1}(k) - \omega_{j_2}(k) = (k-k_c) g(k), 
\end{equation} 
where  $g(k)$ is an analytic function and nonvanishing at $k=k_c$. 
The generalization to a higher-order case is straightforward because we can treat  higher-order poles as  first-order poles by the similar procedure in Eq.~(\ref{eq:second-order-pole}).

Then, $f_{j_1}$ and $f_{j_1}$ can be written as
\begin{align}
f_{j_1}(k)     =  \frac{Q(\omega_{j_1}(k), k)}{p_n \prod_i( \omega_{j_1}(k) - \omega_i(k) )}
    = \frac{1}{k-k_c} \cdot \frac{Q(\omega_{j_1}(k), k)}{p_n g(k) \prod_{i\neq j_1}( \omega_{j_1}(k) - \omega_i(k) )},\\
f_{j_2}(k)     =  \frac{Q(\omega_{j_2}(k), k)}{p_n \prod_i( \omega_{j_2}(k) - \omega_i(k) )}
    = \frac{-1}{k-k_c} \cdot \frac{Q(\omega_{j_2}(k), k)}{p_n g(k) \prod_{i\neq j_2}( \omega_{j_2}(k) - \omega_i(k) )}.
\end{align}
If we introduce the following functional,
\begin{equation}
F(\omega_j(k), k) \equiv  \frac{Q(\omega_{j}(k), k)}{p_n g(k) \prod_{i\neq j}( \omega_{j}(k) - \omega_i(k) )},
\end{equation}
we can write $f_{j_1}(k)$ and $f_{j_2} (k)$ as the form,
\begin{align}
f_{j_1}(k) = \frac{1}{k-k_c} F(\omega_{j_1}(k), k), 
\qquad f_{j_2}(k) = \frac{-1}{k-k_c} F(\omega_{j_2}(k), k).
\end{align}
At $k=k_c$, $F(\omega_1(k_c),k_c)=F(\omega_2(k_c),k_c)$, i.e., the residues of $ f_{j_1}(k)$ and $ f_{j_2}(k)$  have the oposite sign: $\mathrm{Res} f_{j_1}(k_c)= - \mathrm{Res} f_{j_2}(k_c)$.
From this fact, the equation (\ref{eq:G}) in the space-like region turns out to be
\begin{align}
\G( x^\mu ) &= \frac{1}{4\pi r^2} \int^{\infty}_{-\infty} dk \frac{k}{k-k_c} 
 \biggl[ F(\omega_1(k),k) e^{-i\omega_1(k) t+ikr} - F(\omega_2(k), k)e^{-i\omega_2(k t)+ikr}\biggr]\notag \\
   &= \frac{ik_c}{2 r^2}\biggl[ F(\omega_1(k_c),k_c) e^{-i\omega_1(k_c) t+i k_c r} 
        - F(\omega_2(k_c), k_c)e^{-i\omega_2(k_c )t+i k_c r}\biggr] \notag\\
 &=0.
\end{align}
Therefore, the contributions from poles vanish in $\G( x^\mu ) $.

Next, we consider the contribution from Eq.(\ref{eq:pnkpole}).
We suppose that Eq.~(\ref{eq:omegapole}) dose not have any solutions, and $p_n (k) $ is given as 
\begin{align}
p_n (k) = k^2+k_c^2,
\end{align} 
where $k_c$ is a real positive constant.
In this case, the retarded Green function in the space-like region becomes
\begin{align}
G_R(x^\mu ) & = \frac{1}{4\pi r^2} \sum_j \int^{\infty}_{-\infty} dk \frac{k}{k^2+k^2_c} g_j (k) \exp \biggl[ ikr -i \omega_j(k) t\biggr], 
                          \nonumber \\
                  & = \frac{i}{4r^2}\sum_{j} g_j (i k_c) \exp \biggl[ -k_c r -i \omega_j (i k_c) t \biggr], \label{eq:acausal}
\end{align}
where we have supposed that Eqs~(\ref{eq:recondition}) and (\ref{eq:imcondition}) are satisfied, and  introduced 
\begin{align}
g_j (k) \equiv \frac{Q(\omega_j(k), k)}{\prod_{i \neq j} (\omega_j (k) - \omega_i(k))}.
\end{align} 
We can see that Eq.~(\ref{eq:acausal}) does not vanish and  causality is violated.

We note that $p_n (k)$ necessarily yields the poles in the complex $k$ plane if it depends on $k$ 
because  it is the polynomial of $k$.
Therefore, $p_n (k)$, which is the coefficient of the highest time derivative, must not include $k$.

\section{Summary and outlook}\label{Sec:Summary}
We studied relativistic causality in an effective theory based on the derivative expansion.
We first discussed that the Lorentz covariance and the equal order of space and time derivatives in the equation of motion do not ensure causality.

In general, the Green function in the derivative expansion can be generally written as a rational function in the momentum space, Eq.(\ref{eq:Gpoly}).
Using this Green function,  we derived the general conditions ensuring causality, i.e., the Green function vanishes in the space-like region.
The conditions are given by Eqs.~(\ref{eq:recondition}), (\ref{eq:imcondition}) 
and that the coefficient of the highest-order time derivative does not include the space derivative.
Our condition Eq.~(\ref{eq:recondition}) is consistent with the condition about the front velocity.
Furthermore, we obtained the condition for the imaginary parts of poles, Eq.~(\ref{eq:imcondition}).
The diffusion mode Eq.~(\ref{eq:diffusionmode}) satisfies Eq.~(\ref{eq:recondition}), but violate causality due to Eq.~(\ref{eq:imcondition}).

In this paper, we considered the Green function of the scalar fields. Our results will not change for other fields such as vector fields, 
because the analyticity of the Green function is independent of spin or helicity.
It is a remaining task to study the effect of  the thermal and quantum fluctuations which modify the Green function. 
In this case, the Green function has branch cuts and may not be written by a simple rational function Eq.~(\ref{eq:Gpoly}).

 It is discussed that the derivative expansion causes unphysical instability in the context of relativistic hydrodynamics~\cite{Hiscock:1985zz, Li:2010fr, Pu:2009fj, Denicol:2008ha}. It seems that relativistic hydrodynamics satisfying causality conditions are stable~\cite{Denicol:2008ha}.
However,  it is not clear what condition ensures the stability in an effective theory based on the derivative expansion,
which is beyond the scope of this paper, and we leave it for future work.

\acknowledgements
We acknowledge various discussions at RIKEN Open House 2013.  
This work  was partially supported by JSPS KAKENHI Grants Numbers 24740184,  23340067, 
and by RIKEN iTHES Project.

\bibliography{causality}
\end{document}